\documentclass[twocolumn,showpacs,preprintnumbers,amsmath,amssymb]{revtex4}

\usepackage{graphicx}
\usepackage{dcolumn}
\usepackage{bm}
\usepackage{enumerate}

\begin{document}


\title{N-tuple Zipf Analysis and Modeling for Language, Computer Program and DNA}

\author{Xiaocong Gan}
\email{sliant@mail.bnu.edu.cn}
\author{Dahui Wang}
\email{wangdh@bnu.edu.cn}
\author{Zhangang Han}
\email{zhan@bnu.edu.cn}
\affiliation{
Department of Systems Science, Beijing Normal University, Beijing, 100875, P. R. China
}


\begin{abstract}
\emph{n}-tuple power law widely exists in language, computer program code, DNA and music. After a vast amount of Zipf analyses of \emph{n}-tuple power law from empirical data, we propose a model to explain the \emph{n}-tuple power law feature existed in these information translational carriers. Our model is a preferential selection approach inspired by Simon's model which explained scaling law of single symbol in a sequence Zipf analysis. The kernel mechanism is neat and simple in our model. It can be simply described as a randomly copy and paste process, that is, randomly select a random segment from current sequence and attach it to the end repeatedly. The simulation of our model shows that \emph{n}-tuple power law exists in model generated data. Furthermore, two estimation equations: the Zipf exponent and the minimal length of \emph{n}-tuple for power law appears all correspond to empirical data well. Our model can also reproduce the symmetry breaking process of ATGC number differences in DNA data.
\end{abstract}



\maketitle

\section{\label{sec1}Introduction}

An intriguing feature of language is the Zipf's law \cite{Zipf49}, also known as Power Law or Pareto's Law \cite{Adamic00}.
In the Zipf analysis, one calculate the frequencies of each word in an English text, and sort all the frequencies in rank-order, from the largest to the smallest. If we plot these frequencies against their rank-order in a log-log figure, then it will show a nearly straight line, with a slope $\xi \approx -1$ \cite{Zipf49}. So the relation of frequency and corresponding rank-order can be approximated by a power law form, $k = r ^ {\xi}$, $k$ for frequency, $r$ for rank, and $\xi$, usually negative, is referenced as Zipf exponent. Some researches have used an extended form $k = ( r + c ) ^ {-a}$ \cite{Zanette06, Montemurro02, Mandelbrot53, Mandelbrot66}. The constant $c$, however, does not have a substantial physical meaning. \cite{Rio08, Naumis08} showed that instead of a power law, many different data in rank laws can be very well fitted by the integrand of a beta function. Zipf analysis were also extended to many other systems\cite{Mandelbrot83}, such as city sizes \cite{Gell-Mann94}, DNA base pair sequences \cite{Mantegna94}, and distribution of firm sizes \cite{Stanley95}.

Many information carriers, such as language, program code, and DNA, can be considered as a symbol sequence. English text can be regarded as word sequences or letter sequences, where words are distinguished by some letters separated by space or punctuation. Chinese text can be perceived as Chinese character sequences, computer binary file as binary sequences, and DNA as ATGC sequences. It's well known that statistics on single symbol of these sequences show no power law except for English text as word sequence \cite{Kanter95, Zipf49}. For example, statistics on 26 letters in an English novel, on Chinese characters in a Chinese novel, or on the 4 symbols ATGC in a DNA sequence, show no power law.

Words are not easily separated in some languages as they are separated with spaces in English corpus. For example, in Chinese, compound words composing two or more Chinese characters could be created if they are semantically meaningful \cite{Ha03}. Also in other sequences, e.g., noncoding regions of DNA sequences, it is not easy to distinguish words \cite{Mantegna94}. Literature reported that statistics on \emph{n}-tuples in these symbol sequences show Zipf's law \cite{Mantegna94, Ha03}. Let's use an example to demonstrate the statistic method. Given an English letter sequence, ``abbccc'', its length is 6. Then we get 5 2-tuples: ``ab'', ``bb'', ``bc'', ``cc'', ``cc''. There are 4 unique 2-tuples here: ``ab'', ``bb'', ``bc'', ``cc'', with frequencies 1, 1, 1, 2 respectively. Formally, given a symbol sequence $S=(s_{1},s_{2},\cdots,s_{t})$, its length is $t$. Then we get $t-n+1$ \emph{n}-tuples, imaging a window with width $n$ slide from the beginning to the end. We can perform frequency statistics on these \emph{n}-tuples. If the statistic results show power law, then we call it \emph{n-tuple Power Law}. The phrase \emph{n}-tuple used in \cite{Mantegna94, Kanter95, Czirok95, Czirok96} is also called \emph{n}-gram in \cite{Damashek95, Ha03}. We inherit \emph{n}-tuple in this paper.

\cite{Mantegna94} reported that \emph{n}-tuples of DNA (noncoding regions) as ATGC base pair sequence demonstrates a Zipf feature. This feature also exists in \emph{n}-tuples of English text as letter sequence, and computer binary executable file as 0, 1 sequence. In that paper, \emph{n}-tuple Zipf analyses were performed on DNA with $n=3$ through 8, and on English text with $n=3$ through 5, and $n=12$ on computer binary executable file. \cite{Mantegna94} also claimed that noncoding sequences bear more resemblance to a nature language than the coding sequences. \cite{Israeloff96} argued, however, to detect such linguistic feature, Zipf analysis should be applied with caution, since it cannot distinguish language from power-law noise, e.g., $1/f$ noise.

\cite{Ha03} give a detailed report that \emph{n}-tuple power law exists in English text as word or letter sequence, and in Chinese text from 1-tuple to 5-tuple.

However, our statistics show that \emph{n}-tuple power law exists for a much larger $n$ and in ranges different from \cite{Mantegna94}. For human DNA (note that we do not distinguish coding and noncoding regions), when $n\geq10$ , statistics show a better power law. For English text, a better power law shows when $n\geq4$. \cite{Mantegna94} also reported that the Zipf exponent $\xi$ is almost the same for different $n$, which is found to be increasing \cite{Ha03, Israeloff96}. Our statistics also show an increasing Zipf exponent with $n$ increases. In Section~\ref{sec3} we'll give an estimate equation for the range of $n$, based on our model.

\cite{Czirok95, Czirok96, Kanter95} proposed Markov process to analyze the \emph{n}-tuple where the sequence is simplified to contain only two different symbols, 0, and 1. Conditional probability was calculated and gave results roughly similar to the one observed for long-range correlated sequences. \cite{Egghe00} gives the rank-frequency distribution of \emph{n}-tuples based on the assumption that the rank-frequency distribution of single symbols follows Zipf's law.

Simon proposed a preferential selection model to explain the power law distribution in numerous examples with this property found at that age \cite{Simon55}. However, Simon's model cannot explain the \emph{n}-tuple Zipf feature. This paper will follow Simon's idea and set up a model to explain the \emph{n}-tuple Zipf feature.

In Section~\ref{sec2}, we will give our statistics on English corpus, Chinese corpus, DNA, computer program coding source code, and computer executable binary file. We'll show that for a random sequence, a Zipf's law does not exist. In Section~\ref{sec3}, we'll propose a model to explain the \emph{n}-tuple power law. Later, we'll give an estimation equation for the Zipf exponent. We draw conclusions in section 4.

\section{\label{sec2}Empirical results}

\subsection{\label{sec2.1}Zipf analysis of \emph{n}-tuples}

\cite{Montemurro01} mentioned that short horizontal line segments appeared at the bottom of a Zipf plot interfere with the statistics, and proposed that the last one or two of these line segments be discarded and the rest of them be represented by their center point respectively. Here we adopt a similar method: for all the line segments, we preserve the right-most point and discard the rest. This is a much easier way to eliminate the interference.

We do the traditional Zipf plot and then we perform a linear fit on the log-log plot. The slope is the Zipf exponent (negative), and the square of correlation coefficient $\rho^{2}\in[0,1]$ represents how well the fit is, with 1 a perfect straight line and 0 not a line at all. We say it's a power law if $\rho^{2}$ is close to 1 (Typically when $\rho^{2}\geq0.95$).

We perform statistics on English corpus, Chinese corpus, DNA, computer program coding source code, and computer executable binary file. We also perform statistics on DNA as 01 sequence with AT=0 and GC=1, music pieces as music note sequence, and actor sequence in drama. We show here only statistics on English corpus, and DNA sequence. The rest of the statistical results are presented in supplement material \cite{sup}. Because almost all \emph{n}-tuples appear only one or two times when $n$ is too large, we only perform statistics for relatively not too large $n$. Note that it's quite time consuming to perform \emph{n}-tuple statistics on large data sets, e.g. human DNA. State to the art technique is needed. Some programming techniques we used is represented in \cite{sup}.

\begin{figure}[h]
\includegraphics{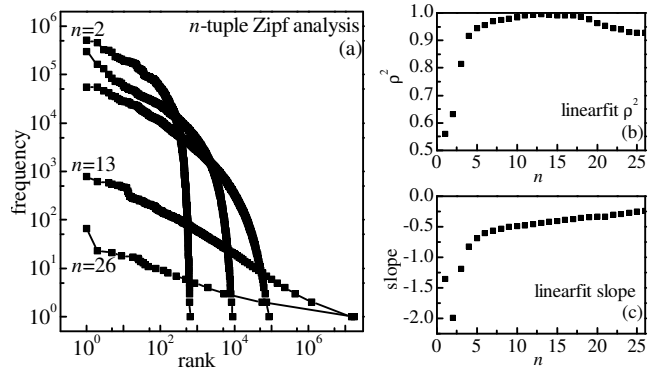}
\caption{\label{fig1}
\emph{n}-tuple Zipf analysis of 15 novels of Dickens as English letter sequence.
(a) The frequency-rank statistics on \emph{n}-tuples for $n=2,3,4,13,26$.
(b) $\rho^{2}$ of linear fit against $n$. We can see that \emph{n}-tuple Zipf analyses show power law for $n\geq4$.
(c) Slope of linear fit against $n$. We can see that the slope tends to zero.
}
\end{figure}

\begin{figure}[h]
\includegraphics{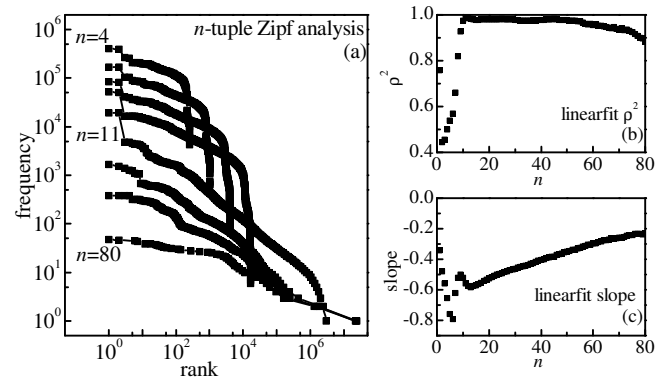}
\caption{\label{fig2}
\emph{n}-tuple Zipf analysis of Y chromosome of human being as ATGC sequence. Source: NCBI Human Genome Resources \cite{dna}.
(a) The frequency-rank statistics on \emph{n}-tuples for $n=4,5,6,7,11,25,40,80$.
(b) $\rho^{2}$ of linear fit against $n$. We can see that \emph{n}-tuple Zipf analyses show power law for $n\geq10$.
(c) Slope of linear fit against $n$. We can see that the slope tends to zero as in Fig.~\ref{fig1}.
}
\end{figure}

Fig.~\ref{fig1} and Fig.~\ref{fig2} are the Zipf analyses of English corpus of Dickens' 15 novels as letter sequence and DNA ATGC sequence of human Y chromosome from \cite{dna}. For Dickens novels, when $n\geq4$, it is already a well fit to a power law. For Y chromosome, however, it is when $n\geq10$.

We can see from Fig.~\ref{fig1}(c) and Fig.~\ref{fig2}(c) that the slope (the Zipf exponent) tends to zero with $n$ increasing. We find that this is the case in all our statistics \cite{sup}. In Section~\ref{sec3} we'll try to explain this feature based on our model.

\subsection{\label{sec2.2}No \emph{n}-tuple power law in random sequence}

\begin{figure}[h]
\includegraphics{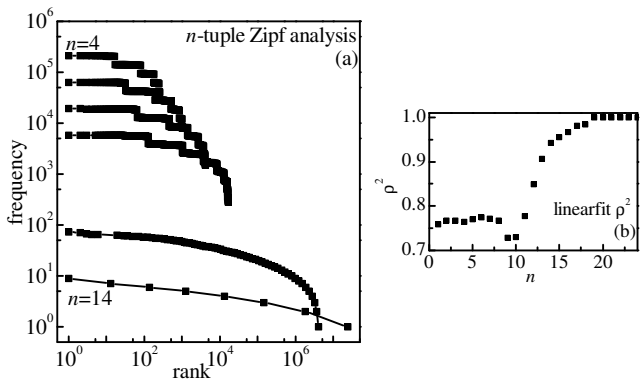}
\caption{\label{fig3}
\emph{n}-tuple Zipf analysis of a random (shuffled) ATGC sequence.
(a) The frequency-rank statistics on \emph{n}-tuples for $n=4,5,6,7,11,14$.
(b) $\rho^{2}$ of linear fit against $n$. We can see that $\rho^{2}$ is close to 1 for $n\geq14$. This is, however, not an evidence for a power law. The reason is that almost all \emph{n}-tuples appear only a few times when $n\geq14$. For example, all \emph{n}-tuples appear only one or two times for $n=19$, so there are only two points on the curve. In this case, $\rho^{2}$ of linear fit needs to be exact 1 according to our statistic method. As mentioned in Section~\ref{sec2.1}, these should be discarded. For the rest part $n<14$, it's not power law because $\rho^{2}$ is way too low. Compare (b) with Fig.~\ref{fig2}(b), the difference is clear: in Fig.~\ref{fig2}(b), for the whole range between $n=10$ to 80, $\rho^{2}$ is close to 1 which indicates power law.
}
\end{figure}

It should be noticed that \emph{n}-tuple power law does not exist in random sequence. We generate a random ATGC sequence, the length of the sequence and the numbers of  A, T, G, and C are the same as the real Y chromosome from the above source. In fact, such a sequence is a shuffle of the original one. Fig.~\ref{fig3} is an \emph{n}-tuple Zipf analysis on such a shuffled sequence. We can see in Fig.~\ref{fig3}(b) that the \emph{n}-tuple curves are not linear when $n<14$ because $\rho^{2}$ is low. Although when $n\geq14$, the curve is linear and $\rho^{2}$ is high (close to 1), this is not evident for a power law. The reason is due to the fact that, in our statistic method, when $n\geq14$, there are only a few points on the curve, exactly speaking, 8 points for $n=14$, and 2 points for $n=19$, in which case the $\rho^{2}$ of linear fit needs to be exact 1.

We can give an explanation of why \emph{n}-tuple power law does not exist in random sequence. Given a random ATGC sequence, suppose each element in the sequence is an independent and identically distributed random variable, and the probability for ATGC is $P_{A},P_{T},P_{G},P_{C}>0$ respectively. Then the probability that two elements at two given loci are the same is $P=P_{A}^{2}+P_{T}^{2}+P_{G}^{2}+P_{C}^{2}$, $0<P<1$. Two \emph{n}-tuples are the same means elements on every corresponding loci are the same, so the probability is $P^{n}$. $P^{n}$ tends to 0 exponentially with $n$ increases. That's why almost all \emph{n}-tuples appear only a few times when $n\geq14$ in Fig.~\ref{fig3}. When $n$ is relatively small, the probability of each unique \emph{n}-tuple could be easily calculated with $P_{A},P_{T},P_{G},P_{C}$, showing no way of being a power law.

\section{\label{sec3}Modeling \emph{n}-tuple}

\subsection{\label{sec3.1}Simon's model}

Simon's model is a preferential selection model \cite{Simon55}. It can be simply described as: randomly select a element from current sequence and attach it to the end of the sequence repeatedly. A formal description is as follows: at each step a new element is attached to the end of current symbol sequence. The newly attached element follows two rules:

\begin{description}
\item[Rule 1 (new unique symbol rule).]
There is a constant probability $\alpha$ that the newly attached element is a new unique symbol that never appeared.
\item[Rule 2 (preferential selection rule).]
Else the newly attached element is randomly selected from the current sequence.
\end{description}

From these two rules, Simon proved that power law will appear and the slope (Zipf exponent) is $\alpha-1$.

Note that although Simon's model is still valid when $\alpha$ is very close to 0 or 1, it is not easy to observe a power law at this circumstance due to the fact that the sequence length needs to be very large to exhibit any meaningful feature.

We've performed \emph{n}-tuple Zipf analysis on Simon's model generated sequence, and found that \emph{n}-tuple power law does not exist. The plot is similar to Fig.~\ref{fig3}. So, we need a new model that can compromise \emph{n}-tuple power law.

\subsection{\label{sec3.2}Model description and simulation results}

Our model is a consecutive subsequence preferential selection model inspired by Simon's model. It can be simply described as a randomly copy and paste process: randomly select a random consecutive subsequence from current sequence and attach it to the end repeatedly. A formal description is as following:

\begin{enumerate}[Step \arabic{enumi}.]
\setcounter{enumi}{-1}
\item
Given 4 parameters: the length $T_{min}$ of initial symbol sequence, the number $U$ of unique symbols, the discrete distribution $D$ which generates random positive integers, and the maximum length $T_{max}$ of symbol sequence.
\item
Generate an initial symbol sequence, in which each element is randomly selected from $U$ unique symbols.
\item
Suppose the current sequence is $C$, and the length is $t$. Generate a random integer $a$, which has a uniform distribution in $[1,t]$, as the start point of subsequence. Generate a random length $b$, which has a distribution $D$, as the length of subsequence. If $a+b\leq t+1$, go to Step 3; else, repeat this step. (This is to make sure the randomly selected subsequence is inside $C$)
\item
Suppose $R$ is the randomly selected subsequence in $C$, starting from $a$ with length $b$. $R$ is copied and attached to the end of $C$ and assign $C$ as the new sequence. Update the length of $C$: $t=t+b$.
\item
If $t>T_{max}$, stop; else, go to Step 2.
\end{enumerate}

\begin{figure}[h]
\includegraphics{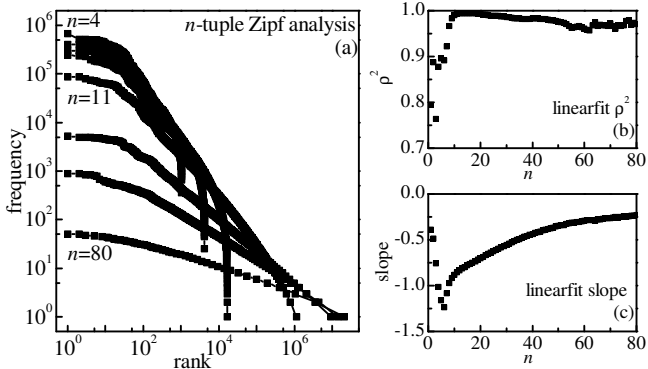}
\caption{\label{fig4}
\emph{n}-tuple Zipf analysis of our model generated data. Two related parameters of our model are set up according to real human Y chromosome as in Fig.~\ref{fig2}: $U=4$, $T_{max}=25652954$. $T_{min}=100$, $D$ is exponential distribution with PDF (Probability Distribution Function) $0.02e^{-0.02x}$ (use the integer part of generated random numbers).
(a) The frequency-rank statistics on \emph{n}-tuples for $n=4,5,6,7,11,25,40,80$.
(b) $\rho^{2}$ of linear fit against $n$. We can see that \emph{n}-tuple Zipf analyses show power law for $n\geq10$.
(c) Slope of linear fit against $n$. We can see that the slope tends to zero.
These are the same as in real DNA data shown in Fig.~\ref{fig2}.
}
\end{figure}

Fig.~\ref{fig4} is an \emph{n}-tuple Zipf analysis on this model generated sequence. Parameters of the model are tuned to real DNA as in Fig.~\ref{fig2}. We can see that \emph{n}-tuple power law does exist in our model and it well replicates the DNA Zipf analysis as in Fig.~\ref{fig2}.

\subsection{\label{sec3.3}Model Analysis}

\begin{figure}[h]
\includegraphics{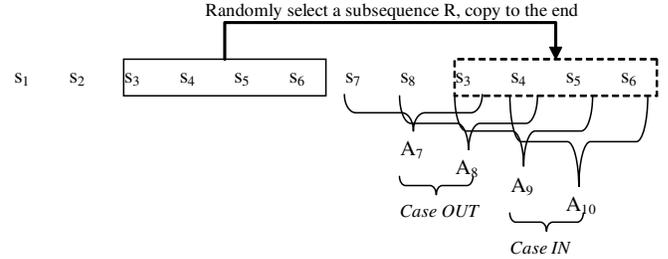}
\caption{\label{fig5}
A demonstration of our model. $R=(s_{3},s_{4},s_{5},s_{6})$ is randomly selected, copied and attached to the end.
}
\end{figure}

Let's begin with an example. Suppose the current sequence is $C=(s_{1},s_{2},\cdots,s_{8})$, length 8. There are 6 3-tuples: $A_{1}=(s_{1},s_{2},s_{3})$, $A_{2}=(s_{2},s_{3},s_{4})$, $\cdots$, $A_{6}=(s_{6},s_{7},s_{8})$. Randomly select a subsequence from $C$, say, starting at 3 with length 4, that is $R=(s_{3},s_{4},s_{5},s_{6})$. Copy and attach $R$ to the end of $C$, now $C=(s_{1},s_{2},\cdots,s_{8},s_{3},s_{4},s_{5},s_{6})$, length 12. There are 10 3-tuples now: $A_{1},A_{2},\cdots,A_{6}$ are the same, and $A_{7}=(s_{7},s_{8},s_{3})$, $A_{8}=(s_{8},s_{3},s_{4})$, $A_{9}=(s_{3},s_{4},s_{5})$, $A_{10}=(s_{4},s_{5},s_{6})$ are newly formed. We can see that  $A_{9}=A_{3}$, $A_{10}=A_4$. As of $A_7$ or $A_8$, it's unknown if it equals to any of $A_{1},A_{2},\cdots,A_{6}$. If it's not, then a new unique 3-tuple is introduced. Fig.~\ref{fig5} shows this whole process.

Now we can give a formal description. Suppose the current sequence is $C=(s_{1},s_{2},\cdots,s_{t})$, length $t$. Let's consider \emph{n}-tuples, there are $t-n+1$ \emph{n}-tuples: $A_{1}=(s_{1},s_{2},\cdots,s_{n})$, $A_{2}=(s_{2},s_{3},\cdots,s_{n+1})$, $\cdots$, $A_{t-n+1}=(s_{t-n+1},s_{t-n+2},\cdots,s_{t})$. A subsequence $R=(s_{a},s_{a+1},\cdots,s_{a+b-1})$ is randomly selected from $C$, starting at $a$ with length $b$, and is copied and attached to the end of $C$. Now $C=(s_{1},s_{2},\cdots,s_{t},s_{a},s_{a+1},\cdots,s_{a+b-1})$, length $t+b$. There are $t-n+1+b$ \emph{n}-tuples now: $A_{1},A_{2},\cdots,A_{t-n+1}$ are the same, $A_{t-n+2},A_{t-n+3},\cdots,A_{t-n+1+b}$ are newly formed. The number of these newly formed is $b$. These newly formed can be divided into two cases, \emph{Case IN} and \emph{Case OUT} as Fig.~\ref{fig5} shows.

\begin{description}
\item[Case IN.]
If $b\geq n$, among these newly formed $b$ \emph{n}-tuples, the last $b-n+1$ \emph{n}-tuples fall inside of $R$, and equal to $A_{a},A_{a+1},\cdots,A_{a+b-n}$ respectively. Suppose $P_{in}$ is the probability that a newly formed \emph{n}-tuples belongs to this case.
\item[Case OUT.]
If $b\geq n$, among these newly formed $b$ \emph{n}-tuples, the first $n-1$ \emph{n}-tuples fall (partly) outside of $R$. If $b<\lambda$, all the newly formed $b$ \emph{n}-tuples fall (partly) outside of $R$. It's unknown whether these newly formed \emph{n}-tuples equal to any of $A_{1},A_{2},\cdots,A_{t-n+1}$. If it's not, then a new unique \emph{n}-tuple is introduced.
\end{description}

Unfortunately, we are unable to give a strict analysis for \emph{Case OUT}. Therefore we give the following assumption.

\begin{description}
\item[Assumption 1.]
The probability that an element in \emph{Case OUT} is a duplicated one is very small and can be neglected when $n$ is large.
\end{description}

This assumption is necessary for the following discussion. One may doubt the reasonableness of this assumption. Well, the most convincible evidence could be that our theoretical equation based on this assumption fit the model well, see Section~\ref{sec3.4} and Section~\ref{sec3.5}. The fact that the number of all possible \emph{n}-tuples increases exponentially with $n$ increases also favors this assumption, see Section~\ref{sec3.5}. We hope that future work can give a strict analysis for this assumption.

Starting from this assumption, if we perceive $A_1, A_2, \cdots$ in our model as the symbols in Simon's model, \emph{Case IN} is equivalent to the preferential selection rule in Simon's model (Rule 2), and \emph{Case OUT} can correspond to the new unique symbol rule in Simon's model (Rule 1). Now we can utilize the proof of Simon's model and prove the existence of \emph{n}-tuple power law in our model. We can also calculate Zipf exponent. In the next section, we show that the calculated Zipf exponent corresponds well to the model. This demonstrates the validity of this assumption.

\subsection{\label{sec3.4}Slope (Zipf exponent)}

Now we give an estimate for the slope (Zipf exponent) of our model.

Consider $P_{in}$. Given that the PDF of the distribution $D$ is $f$. According to step 2 in Section~\ref{sec3.2}, we need to repeatedly generate random integer $b$ by distribution $D$ until $a+b\leq t+1$. However, we suppose that the generated random integer $b$ always satisfies $a+b\leq t+1$. This is reasonable because this is almost the case for any PDF when $t$ is large. Furthermore, assume that the expectation corresponding to the distribution $D$ exists.

Now, according to \emph{Case IN}, we have
\[P_{in}=\frac{\sum_{b=n}^{+\infty}(b-n+1)f(b)}{\sum_{b=0}^{+\infty}b f(b)},\]
or in the integral form
\[P_{in}=\frac{\int_n^{+\infty}(x-n+1)f(x)\text{d}x}{\int_0^{+\infty}x f(x)\text{d}x}.\]
We use $P_{in}$ to estimate the slope (Zipf exponent). According to \emph{Assumption 1} and Simon's model, $P_{in}=1-\alpha$. So we have
\begin{equation}\label{eq1}
\text{slope}=\alpha-1=-P_{in}=\frac{-\int_n^{+\infty}(x-n+1)f(x)\text{d}x}{\int_0^{+\infty}x f(x)\text{d}x}.
\end{equation}

\begin{figure}[h]
\includegraphics{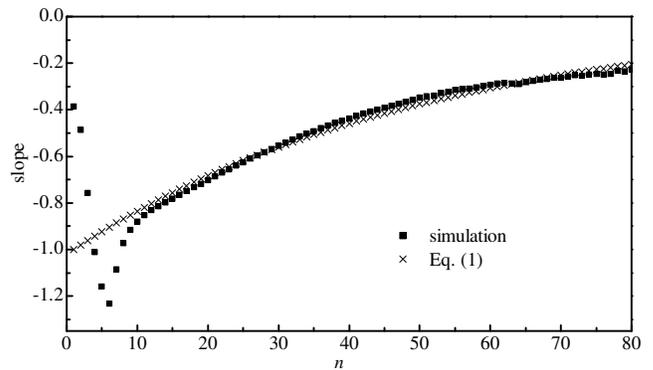}
\caption{\label{fig6}
Compare Eq.~(\ref{eq1}) with model simulation result as in Fig.~\ref{fig4}(c). The parameters are the same as in Fig.~\ref{fig4}. We can see that Eq.~(\ref{eq1}) fits the model well when $n$ is large ($n\geq10$), which indicate the validity of \emph{assumption 1}.
}
\end{figure}

Eq.~(\ref{eq1}) is got with no requirement for the detailed form of the distribution function. The deduction only requires the existence of the expectation. Fig.~\ref{fig6} is a comparison of Eq.~(\ref{eq1}) and actual simulation result of our model. We can see that when $n$ is large ($n\geq10$), Eq.~(\ref{eq1}) gives the same result as our model gives. This demonstrates the validity of \emph{assumption 1}.

We now compute the limit of Eq.~(\ref{eq1}). Because $f$ is the PDF of distribution $D$ which generates random positive integers, $\int_0^{+\infty}f(x)\text{d}x$ is absolute convergent, so $\lim_{n\to+\infty}\int_n^{+\infty}f(x)\text{d}x=0$. According to the assumption that the expectation corresponding to $D$ exists, $\int_0^{+\infty}x f(x)\text{d}x$ is absolute convergent, so $\lim_{n\to+\infty}\int_n^{+\infty}x f(x)\text{d}x=0$. Note that $0\leq\int_n^{+\infty}n f(x)\text{d}x\leq\int_n^{+\infty}x f(x)\text{d}x$, so $\lim_{n\to+\infty}\int_n^{+\infty}n f(x)\text{d}x=0$. From these, we can extract the numerator of Eq.~(\ref{eq1}) to 3 parts and have
\[\lim_{n\to+\infty}\text{slope}
=
\lim_{n\to+\infty}
    \frac{-\int_n^{+\infty}(x-n+1)f(x)\text{d}x}{\int_0^{+\infty}x f(x)\text{d}x}
=0.
\]

Secondly, let's see the derivative of Eq.~(\ref{eq1}). Notice that the numerator is an integral depending on parameters, we have
\[
\frac{\text{d}}{\text{d}n}\left(\int_n^{+\infty}(x-n+1)f(x)\text{d}x\right)
=
-\int_n^{+\infty}f(x)\text{d}x-f(n)
\leq0,
\]
hence
\[
\frac{\text{d}}{\text{d}n}\left(\text{slope}\right)
=
\frac{-1}{\int_0^{+\infty}x f(x)\text{d}x}
\frac{\text{d}\left(\int_n^{+\infty}(x-n+1)f(x)\text{d}x\right)}{\text{d}n}
\geq0.
\]

The fact that the limit of the slope is zero, and the derivative is greater or equal to zero explains why all the slopes in our statistics and simulations are monotonically increasing and tend to zero when $n$ is large. Notice that when $n$ is small, the slopes in our statistics and simulations are irregular. This is because \emph{Assumption 1} is not satisfied.

\subsection{\label{sec3.5}The threshold of $n$}

We found from the statistics and the simulation of our model that \emph{n}-tuple power law doesn't exist when $n$ is small. We need an estimation of the threshold of $n$ that \emph{n}-tuple power law appears when $n$ is greater than the threshold.

Following notations in the above sections, there are only $U^n$ possible unique \emph{n}-tuples. Because $\alpha$ is the probability that a newly generated \emph{n}-tuple is a new one that has never appeared before, the expected number of unique \emph{n}-tuples in the sequence $C$ with length $t$ is $t$. So we have $\alpha t\leq U^n$, i.e.
\begin{equation}\label{eq2}
n \geq \log _U t + \log _U \alpha.
\end{equation}

We can easily infer the followings.

For any given $n$, we can have a sufficiently large $t$ so that Eq.~(\ref{eq2}) fails to hold. Intuitively, for a given $U$ and a given $n$, when the sequence length increases, the probability that a newly generated \emph{n}-tuple did not occur before is decreasing, instead of being approximately a constant. This is going to be further addressed in the conclusion section.

For any given $t$, we can have a sufficiently small $n$ that Eq.~(\ref{eq2}) fails to hold; in other words, $n$ need to be sufficiently large so that Eq.~(\ref{eq2}) can hold.

Therefore, we use Eq.~(\ref{eq2}) to estimate the threshold of $n$. In all the data we analyze in this paper, $t$ is very large and $\alpha$ is not close to zero, so we have a simpler estimation form
\begin{equation}\label{eq3}
n \geq \log _U t.
\end{equation}

Let's compare Eq.~(\ref{eq3}) with actual statistic results. In Fig.~\ref{fig1}, we perform statistics on 15 novels of Dickens, as English letter sequence. There are totally 17211736 letters, and 26 possible unique letters (a-z), $\log_{26} 17211736=5.113753$. We can see in Fig.~\ref{fig1} that \emph{n}-tuple power law exists for about $n\geq4$. In Fig.~\ref{fig2}, the DNA sequence length is 25652954, with 4 possible unique symbols (ATGC), $\log_4 25652954=12.306311$, and we can see that \emph{n}-tuple power law exists for about $n\geq10$.

This is an interesting result. It reveals the relation between Power Law and diversity. As mentioned above, in order to show Power Law, the number of unique elements in a sequence should not be too small or too large, i.e. a proper diversity should be maintained.

A subtle problem should arouse some attention here. If Simon's model is a sufficient and necessary condition for a power law curve, then violating the above inequality means violating Simon's model hence there is no power law. Unfortunately, Simon's model is only a sufficient condition for power law, not  a necessary one. This means this section is not a strict theoretical deduction. We hope future work can improve this.

\subsection{\label{sec3.6}Model parameter settings}

There are 4 parameters in our model, as mentioned in Section~\ref{sec3.2}, the initial length $T_{min}$, the number $U$ of unique symbols, the discrete distribution $D$, and the maximum length $T_{max}$.

It's obvious that our model requires $1<T_{min}\ll T_{max}$ and $1<U\ll T_{max}$. We did some sensitivity analyses and find that our model is not sensitive to $T_{min}$, $U$ and $T_{max}$. While other parameters remain the same, we test different values for these 3 parameters, for example, $T_{min}=100$, $T_{min}=1000$, $U=4$, $U=100$, $T_{max}=10^6$, $T_{max}=10^7$, etc., and find that \emph{n}-tuple power law always exist in the simulations, with only the threshold of $n$ varies a little according to the discussions of the above section.

However, there are certain requirements for the distribution $D$. As mentioned in Section~\ref{sec3.1}, in order to observe a power law distribution, the probability $\alpha$ that a new unique element is introduced, should not be very close to 0. According to Eq.~(\ref{eq1}), this requires that $\int_n^{+\infty}(x-n+1)f(x)\text{d}x$ is not very close to 0 when $n$ is large, i.e. $P(X>n)$ is not very close to 0 when $n$ is large. Such a distribution can be a distribution corresponding to a large expectation (e.g. an exponential distribution with expectation 50 as in Fig.~\ref{fig4} ), a distribution that has a fat tail, or even a degenerate distribution that has only one large value.

\subsection{\label{sec3.7}The symmetry breaking process}

The number of each ATGC in DNA sequence is not the same. We calculated the entropy of human DNA ATGC data \cite{dna}. The entropy is defined as $-\sum p_i \log _2 p_i$, where $p_i$ is the portion of ATGC in the sequence, $i=1,2,3,4$. The entropy ranges from 1.959566 for chromosome 4 to 1.999227 for chromosome 19 with an average of 1.97733 and a standard deviation of 0.01084.

This stylized fact is reproduced with our model. ATGC in our model initially follows uniform distribution. We calculate the entropy for 1000 simulation runs of our model, with the same parameters for that in Fig.~\ref{fig4}. We find that after the growth process, the entropy of our model ranges from 1.74984 to 1.99991 with an average of 1.98894 and a standard deviation of 0.01652. The symmetry breaking process is due to the selection at the early steps.

\section{\label{sec4}Conclusion}

We do a lot of \emph{n}-tuple Zipf analyses to a very large $n$ in a wide variety of real data ranging covering English corpus, Chinese corpus, computer program source code and binary file, music, and ATGC sequence from human DNA (see supplement material \cite{sup}). These analyses showed the trend that when $n$ increases to a certain value, there exists a power law for sure and the slope tends to zero. We also showed that there is no \emph{n}-tuple power law for random data by reshuffling the DNA data.

We perceive this \emph{n}-tuple scaling law feature in a variety of information carriers as that a meaningful ``motif'' in information translational carrier in each field needs a certain length of symbols to express a relatively complete ``sentence'', hence a motif has a specific characteristic sequence length.

Instead of modeling this \emph{n}-tuple power law feature in a 1-bit Markov process \cite{Czirok95, Czirok96, Kanter95}, we set up a model to explain the \emph{n}-tuple power law feature. The model is a simple copy and paste process, inspired by Simon's model. The model is tuned to DNA data and it showed its effectiveness in reproducing the \emph{n}-tuple power law feature. We hope this model could help to figure out how language and DNA are generated.

Based on \emph{Assumption 1} that \emph{Case OUT} are almost all new unique \emph{n}-tuples, we also calculated the Zipf exponent and proved that it tends to zero, the same trend as real data shows. However, we hope future work can give a strict analysis for \emph{Case OUT} and \emph{Assumption 1}. Moreover, when the length of the sequence increases with our growth mechanism, the probability that \emph{Case OUT} is a new element decreases. \cite{Mandelbrot59, Zanette05} discussed Simon's model under this circumstance.

This model also reproduces the symmetry breaking process of ATGC inequality in DNA sequence with an average entropy value quite close to the real one.

We should point out here that real data have some aspects that this model does not always well address. We do a lot of analyses base on DNA data. We do not calibrate this model to other data sources due to the consideration that analysis based on DNA data already gives the main results of this model. Calibrate this model to other data sources, which is quite time consuming, may not show something new. There are other features in empirical data. For example, in English corpus as letter sequences and Chinese modern corpus as Chinese character sequences, the arrival of a given symbol is a Poisson process, while this is not the case for English corpus as word sequences , Chinese ancient corpus and DNA sequence. This model generated sequences, however, are always Poisson processes (see supplement material \cite{sup}). We also find that long range correlation, which is found in noncoding region of real DNA \cite{Peng92}, does not exist in this model generated sequences.

\begin{acknowledgments}
This research is supported by National Natural Science Foundation of China under grant number 60774085.
\end{acknowledgments}


\begin{thebibliography}{26}
\expandafter\ifx\csname natexlab\endcsname\relax\def\natexlab#1{#1}\fi
\expandafter\ifx\csname bibnamefont\endcsname\relax
  \def\bibnamefont#1{#1}\fi
\expandafter\ifx\csname bibfnamefont\endcsname\relax
  \def\bibfnamefont#1{#1}\fi
\expandafter\ifx\csname citenamefont\endcsname\relax
  \def\citenamefont#1{#1}\fi
\expandafter\ifx\csname url\endcsname\relax
  \def\url#1{\texttt{#1}}\fi
\expandafter\ifx\csname urlprefix\endcsname\relax\def\urlprefix{URL }\fi
\providecommand{\bibinfo}[2]{#2}
\providecommand{\eprint}[2][]{\url{#2}}

\bibitem[{\citenamefont{Zipf}(1949)}]{Zipf49}
\bibinfo{author}{\bibfnamefont{G.~K.} \bibnamefont{Zipf}},
  \emph{\bibinfo{title}{Human Behavior and the Principle of least Effort}}
  (\bibinfo{publisher}{Addison-Wesley}, \bibinfo{address}{Cambridge, MA},
  \bibinfo{year}{1949}).

\bibitem[{\citenamefont{Adamic}(2000)}]{Adamic00}
\bibinfo{author}{\bibfnamefont{L.~A.} \bibnamefont{Adamic}},
  \emph{\bibinfo{title}{Zipf, power-laws, and pareto - a ranking tutorial}}
  (\bibinfo{year}{2000}),
  \urlprefix\url{http://www.hpl.hp.com/research/idl/papers/ranking/}.

\bibitem[{\citenamefont{Zanette}(2006)}]{Zanette06}
\bibinfo{author}{\bibfnamefont{D.~H.} \bibnamefont{Zanette}},
  \bibinfo{journal}{Musicae Scientiae} \textbf{\bibinfo{volume}{10}}
  (\bibinfo{year}{2006}).

\bibitem[{\citenamefont{Montemurro}(2002)}]{Montemurro02}
\bibinfo{author}{\bibfnamefont{M.~A.} \bibnamefont{Montemurro}},
  \bibinfo{journal}{Glottometrics} \textbf{\bibinfo{volume}{4}},
  \bibinfo{pages}{87} (\bibinfo{year}{2002}).

\bibitem[{\citenamefont{Mandelbrot}(1953)}]{Mandelbrot53}
\bibinfo{author}{\bibfnamefont{B.}~\bibnamefont{Mandelbrot}}, in
  \emph{\bibinfo{booktitle}{Communication theory}}, edited by
  \bibinfo{editor}{\bibfnamefont{W.}~\bibnamefont{Jackson}}
  (\bibinfo{publisher}{Butterworths}, \bibinfo{address}{London},
  \bibinfo{year}{1953}), p. \bibinfo{pages}{486}.

\bibitem[{\citenamefont{Mandelbrot}(1966)}]{Mandelbrot66}
\bibinfo{author}{\bibfnamefont{B.}~\bibnamefont{Mandelbrot}}, in
  \emph{\bibinfo{booktitle}{Readings in Mathematical Social Sciences}}, edited
  by \bibinfo{editor}{\bibfnamefont{N.~H.} \bibnamefont{P.F.~Lazarsfield}}
  (\bibinfo{publisher}{Science Research Associates}, \bibinfo{year}{1966}), pp.
  \bibinfo{pages}{151--168}.

\bibitem[{\citenamefont{del Rio}(2008)}]{Rio08}
\bibinfo{author}{\bibfnamefont{M.~B.} \bibnamefont{del Rio}},
  \bibinfo{journal}{Physica A} \textbf{\bibinfo{volume}{387}},
  \bibinfo{pages}{5552} (\bibinfo{year}{2008}).

\bibitem[{\citenamefont{Naumis}(2008)}]{Naumis08}
\bibinfo{author}{\bibfnamefont{G.}~\bibnamefont{Naumis}},
  \bibinfo{journal}{Physica A} \textbf{\bibinfo{volume}{387}},
  \bibinfo{pages}{84} (\bibinfo{year}{2008}).

\bibitem[{\citenamefont{Mandelbrot}(1983)}]{Mandelbrot83}
\bibinfo{author}{\bibfnamefont{B.}~\bibnamefont{Mandelbrot}},
  \emph{\bibinfo{title}{The Fractal Geometry of Nature}}
  (\bibinfo{publisher}{Freeman}, \bibinfo{address}{New York},
  \bibinfo{year}{1983}).

\bibitem[{\citenamefont{Gell-Mann}(1994)}]{Gell-Mann94}
\bibinfo{author}{\bibfnamefont{M.}~\bibnamefont{Gell-Mann}},
  \emph{\bibinfo{title}{The Quark and the Jaguar}}
  (\bibinfo{publisher}{Freeman}, \bibinfo{address}{New York},
  \bibinfo{year}{1994}).

\bibitem[{\citenamefont{Mantegna et~al.}(1994)\citenamefont{Mantegna, Buldyrev,
  Goldberger, Havlin, Peng, Simons, and Stanley}}]{Mantegna94}
\bibinfo{author}{\bibfnamefont{R.~N.} \bibnamefont{Mantegna}},
  \bibinfo{author}{\bibfnamefont{S.~V.} \bibnamefont{Buldyrev}},
  \bibinfo{author}{\bibfnamefont{A.~L.} \bibnamefont{Goldberger}},
  \bibinfo{author}{\bibfnamefont{S.}~\bibnamefont{Havlin}},
  \bibinfo{author}{\bibfnamefont{C.~K.} \bibnamefont{Peng}},
  \bibinfo{author}{\bibfnamefont{M.}~\bibnamefont{Simons}}, \bibnamefont{and}
  \bibinfo{author}{\bibfnamefont{H.~E.} \bibnamefont{Stanley}},
  \bibinfo{journal}{Physical Review Letters} \textbf{\bibinfo{volume}{73}},
  \bibinfo{pages}{3169} (\bibinfo{year}{1994}).

\bibitem[{\citenamefont{Stanley}(1995)}]{Stanley95}
\bibinfo{author}{\bibfnamefont{M.}~\bibnamefont{Stanley}},
  \bibinfo{journal}{Economics Letters} \textbf{\bibinfo{volume}{49}}
  (\bibinfo{year}{1995}).

\bibitem[{\citenamefont{Kanter and Kessler}(1995)}]{Kanter95}
\bibinfo{author}{\bibfnamefont{I.}~\bibnamefont{Kanter}} \bibnamefont{and}
  \bibinfo{author}{\bibfnamefont{D.~A.} \bibnamefont{Kessler}},
  \bibinfo{journal}{Physical Review Letters} \textbf{\bibinfo{volume}{74}},
  \bibinfo{pages}{4559} (\bibinfo{year}{1995}).

\bibitem[{\citenamefont{Ha}(2003)}]{Ha03}
\bibinfo{author}{\bibfnamefont{L.~Q.} \bibnamefont{Ha}}, in
  \emph{\bibinfo{booktitle}{Computational Linguistics and Chinese Language
  Processing,}} (\bibinfo{year}{2003}).

\bibitem[{\citenamefont{Czirok et~al.}(1995)\citenamefont{Czirok, Mantegna,
  Havlin, and Stanley}}]{Czirok95}
\bibinfo{author}{\bibfnamefont{A.}~\bibnamefont{Czirok}},
  \bibinfo{author}{\bibfnamefont{R.~N.} \bibnamefont{Mantegna}},
  \bibinfo{author}{\bibfnamefont{S.}~\bibnamefont{Havlin}}, \bibnamefont{and}
  \bibinfo{author}{\bibfnamefont{H.~E.} \bibnamefont{Stanley}},
  \bibinfo{journal}{Physical Review E} \textbf{\bibinfo{volume}{52}},
  \bibinfo{pages}{446} (\bibinfo{year}{1995}).

\bibitem[{\citenamefont{Czirok et~al.}(1996)\citenamefont{Czirok, Stanley, and
  Vicsek}}]{Czirok96}
\bibinfo{author}{\bibfnamefont{A.}~\bibnamefont{Czirok}},
  \bibinfo{author}{\bibfnamefont{H.~E.} \bibnamefont{Stanley}},
  \bibnamefont{and} \bibinfo{author}{\bibfnamefont{T.}~\bibnamefont{Vicsek}},
  \bibinfo{journal}{Physical Review E} \textbf{\bibinfo{volume}{53}},
  \bibinfo{pages}{6371 } (\bibinfo{year}{1996}).

\bibitem[{\citenamefont{Damashek}(1995)}]{Damashek95}
\bibinfo{author}{\bibfnamefont{M.}~\bibnamefont{Damashek}},
  \bibinfo{journal}{Science} \textbf{\bibinfo{volume}{267}},
  \bibinfo{pages}{843} (\bibinfo{year}{1995}).

\bibitem[{\citenamefont{Israeloff et~al.}(1996)\citenamefont{Israeloff,
  Kagalenko, and Chan}}]{Israeloff96}
\bibinfo{author}{\bibfnamefont{N.~E.} \bibnamefont{Israeloff}},
  \bibinfo{author}{\bibfnamefont{M.}~\bibnamefont{Kagalenko}},
  \bibnamefont{and} \bibinfo{author}{\bibfnamefont{K.}~\bibnamefont{Chan}},
  \bibinfo{journal}{Physical Review Letters} \textbf{\bibinfo{volume}{76}},
  \bibinfo{pages}{1976} (\bibinfo{year}{1996}).

\bibitem[{\citenamefont{Egghe}(2000)}]{Egghe00}
\bibinfo{author}{\bibfnamefont{L.}~\bibnamefont{Egghe}},
  \bibinfo{journal}{Scientometrics} \textbf{\bibinfo{volume}{47}},
  \bibinfo{pages}{237} (\bibinfo{year}{2000}).

\bibitem[{\citenamefont{Simon}(1955)}]{Simon55}
\bibinfo{author}{\bibfnamefont{H.~A.} \bibnamefont{Simon}},
  \bibinfo{journal}{Biometrika} \textbf{\bibinfo{volume}{42}}
  (\bibinfo{year}{1955}).

\bibitem[{\citenamefont{Montemurro}(2001)}]{Montemurro01}
\bibinfo{author}{\bibfnamefont{M.~A.} \bibnamefont{Montemurro}},
  \bibinfo{journal}{Physica A} \textbf{\bibinfo{volume}{300}}
  (\bibinfo{year}{2001}).

\bibitem[{sup()}]{sup}
\bibinfo{note}{the supplement material is in the source of this paper at arXiv \url{http://arxiv.org/abs/0908.0500}}

\bibitem[{dna()}]{dna}
\bibinfo{note}{NCBI Human Genome Resources, Dec. 17, 2008},
  \urlprefix\url{http://www.hpl.hp.com/research/idl/papers/ranking/}.

\bibitem[{\citenamefont{Mandelbrot}(1959)}]{Mandelbrot59}
\bibinfo{author}{\bibfnamefont{B.}~\bibnamefont{Mandelbrot}},
  \bibinfo{journal}{Information And Control} \textbf{\bibinfo{volume}{2}}
  (\bibinfo{year}{1959}).

\bibitem[{\citenamefont{Zanette}(2005)}]{Zanette05}
\bibinfo{author}{\bibfnamefont{D.~H.} \bibnamefont{Zanette}},
  \bibinfo{journal}{Journal of Quantitative Linguistics}
  \textbf{\bibinfo{volume}{12}} (\bibinfo{year}{2005}).

\bibitem[{\citenamefont{Peng}(1992)}]{Peng92}
\bibinfo{author}{\bibfnamefont{C.~K.} \bibnamefont{Peng}},
  \bibinfo{journal}{Nature} \textbf{\bibinfo{volume}{356}},
  \bibinfo{pages}{168} (\bibinfo{year}{1992}).

\end{thebibliography}
\end{document}